\documentclass{article}
\usepackage{stmaryrd}
\usepackage{amsmath}
\usepackage{amsthm}
\usepackage{amssymb}

\usepackage{natbib}
\usepackage{imakeidx}

\date{}

\newcommand{\pp}{\hangindent=16pt\textsc{Exp:}\space}
\newcommand{\qq}{\hangindent=16pt\textsc{Int:}\space}
\newcommand{\ppp}{\hangindent=16pt\hspace{13pt}\space}

\newcommand{\pppp}{\hangindent=16pt\textsc{Expositor:}\space}
\newcommand{\qqqq}{\hangindent=16pt\textsc{Interlocutor:}\space}

\parskip=\medskipamount

\begin{document}

\title{Conversations on Contextuality}

\author{Ehtibar N. Dzhafarov \\
	Purdue University \\
	\and 
	Janne V. Kujala  \\
	University of Jyv\"askyl\"a \\
	}

\maketitle

\subsection*{Dramatis person\ae: }
\begin{description}
\item [\textnormal{\textsc{Expositor,}}] trying to present and clarify the main points
of a certain view of contextuality.
\item [\textnormal{\textsc{Interlocutor,}}] skeptical but constructive.
\item [\textnormal{\textsc{Authors,}}] supportive of Expositor but sympathetic to
Interlocutor (remain off-stage except for occasionally inserting footnotes). 
\end{description}

\section*{Conversation 1}

\noindent \pppp My dear Interlocutor, as we have agreed, we will
discuss a certain approach to probabilistic contextuality. Its authors
call it, perhaps not too descriptively, the \index{Contextuality-by-Default}Contextuality-by-Default
theory (CbD).\footnote{\textsc{Aut:} The term indeed may not have been optimally chosen.
It may suggest that all systems of measurements are contextual until
proven otherwise. This is not true. The ``by-default'' in the name
of the theory refers to the identification of the measurements as
random variables: their identity always (``by default'') depends
on context.} I think I should begin by giving you an informal overview of the
main ideas. 

\noindent \qqqq My dear Expositor, I always find an informal presentation
of ideas a dubious exercise. If I do not understand the presentation
clearly (which happens often), it is never clear to me whether this
is because it was dumbed down so much as to become deficient of information,
or because the ideas themselves are deficient. Nevertheless I should
let you proceed.

\noindent \pp Let me try. \index{measurement!--- object}\emph{Objects} (or things, or properties
--- choose what you like) are measured under varying conditions, called
\index{measurement!--- context}\emph{contexts}. The measurements are generally random variables\index{measurement!--- as random variable},
and their \index{measurement!---, identity of}identity is defined by \emph{what} is measured (object)
and \emph{under what conditions} it is measured (context). As a result,
the same object measured in different contexts is represented by different
random variables: it is meaningless to ask \emph{why} they are different
(hence the designation ``contextuality-by-default''). Moreover,
measurements made in different contexts, whether of the same object
or of different objects, \index{joint distribution!--- not defined}\emph{do not have a joint distribution}.
One cannot, e.g., speak of their correlation or of the probability
with which they have the same value. All measurements made within
one and the same context, however, are \index{joint distribution!--- within a context}\emph{jointly distributed}.
The overall picture we have therefore is one of stochastically unrelated
to each other islands of jointly distributed measurements (``\emph{bunches}
of random variables''). Is this sufficiently clear? 

\noindent \qq I am not sure. How does one define ``objects'' and
``contexts''? 

\noindent \pp Primitives of a theory cannot be explained conceptually
except in their relations to other primitives of the theory, and their
operational meaning may be outside the theory. The \index{measurement!--- object}``objects'' and
\index{measurement!--- context}``contexts'' are such primitives: formally, they are no more than
labels defining the \index{measurement!---, identity of}identity of a measurement (so that each measurement
is defined by two labels, one for ``what'' and another for ``in
what context'').

\noindent \qq Perhaps we could clarify this with examples. 

\noindent \pp Here is an example. Suppose we pose two Yes/No questions
to randomly chosen people and record their responses. The questions
can be asked verbally or presented in writing. Intuitively, a question
asked is the \index{measurement!--- object}``object'' being measured, the presentation mode (verbal
or written) is \index{measurement!--- context}context, and the response to a given question by a
randomly chosen person is \index{measurement!--- as random variable}measurement, a random variable labeled by
the question asked and by its presentation mode. 

\noindent \qq So we have four random variables, if I understood you
correctly: response to question $A$ presented verbally, response
to question $A$ presented in writing, and analogously for the second
question, $B$. 

\noindent \pp Yes. Let me suggest notation for these four random
variables: 
\[
R_{A}^{V},R_{A}^{W},R_{B}^{V},R_{B}^{W},
\]
or, better still,
\[
\left(R_{A}^{V},R_{B}^{V}\right),\left(R_{A}^{W},R_{B}^{W}\right).
\]
$R$ stands for response (whose values can be Yes or No), the subscript
shows the object (question), and the superscript shows the context
(presentation mode). We record $R_{A}^{V}$ and $R_{B}^{V}$ together
because the two questions are posed to a same person. For the same
reason, $R_{A}^{W}$ and $R_{B}^{W}$ are recorded together. Therefore
$\left(R_{A}^{V},R_{B}^{V}\right)$ are jointly distributed, and so
are $\left(R_{A}^{W},R_{B}^{W}\right)$. This accords with the general
rule: random variables recorded in the same context are jointly distributed.
Now, we assume that a person is asked either two written questions
or two verbal questions. So, $R_{A}^{V}$ is never recorded together
with (which means here, never obtained from the same person as) $R_{A}^{W}$
or $R_{B}^{W}$. Therefore such pairs as $R_{A}^{V},R_{A}^{W}$ or
$R_{A}^{V},R_{B}^{W}$ have no joint distribution.

\noindent \qq What does this mean exactly, not to have a joint distribution\index{joint distribution!--- not defined}?

\noindent \pp It is clear that for each of our four variables we
have well-defined probabilities with which their value is Yes: this
is the probability with which a randomly chosen person will respond
Yes to the corresponding question in the corresponding context. Say,
the probability of the event $\left[R_{A}^{V}=\textnormal{Yes}\right]$
is 0.4, the probability of $\left[R_{B}^{V}=\textnormal{Yes}\right]$
is 0.5, and for $\left[R_{B}^{W}=\textnormal{Yes}\right]$ the probability
is, say, 0.7. All these probabilities are well-defined theoretically
and can be estimated empirically. Since we ask $A$ and $B$ in the
verbal mode together (from a same person), we can also define and
estimate the joint probability of $\left[R_{A}^{V}=\textnormal{Yes and }R_{B}^{V}=\textnormal{Yes}\right]$.
For instance, if it equals $0.4\times0.5$, then the two random variables
are independent; if it equals 0.4, they are maximally positively correlated,
etc.

\noindent \qq I see. And, I understand, the situation is different
with $R_{A}^{V},R_{B}^{W}$ or $R_{B}^{V},R_{B}^{W}$, because there
is no meaning in which one could define ``and'' in, say, $\left[R_{B}^{V}=\textnormal{Yes and }R_{B}^{W}=\textnormal{Yes}\right]$.

\noindent \pp Precisely. The probability of this conjunction is undefined
and cannot be estimated empirically. So the general rule is: no two
random variables recorded in different (mutually exclusive) contexts
possess a joint distribution. Let's agree to call such variables \index{stochastically
unrelated}\emph{stochastically
unrelated} (to each other).

\noindent \qq What if I modified the design of the survey, and asked
only one question per person, in writing or verbally? Wouldn't then
even $R_{A}^{V}$ and $R_{B}^{V}$ be stochastically unrelated? And
wouldn't this contradict our general rules?

\noindent \pp No combinations of objects and contexts can contradict
our general rules because these combinations have to be chosen in
accordance with these rules. In your modified set-up, if we continue
to view the questions $A$ and $B$ as our sole and distinct objects,
then the contexts involve not only the mode of presentation but also
the identity of the questions themselves: $\left(A,V\right)$, $\left(B,W\right)$,
etc. So the random variables we record are
\[
R_{A}^{\left(A,V\right)},R_{A}^{\left(A,W\right)},R_{B}^{\left(B,V\right)},R_{B}^{\left(B,W\right)}.
\]
Since no two of them share a context, they are pairwise stochastically
unrelated. 

\noindent \qq But how do I know which of the representations to use,
$R_{A}^{V}$ or $R_{A}^{\left(A,V\right)}$?

\noindent \pp First you have to decide (outside the CbD theory) on
the empirical meaning of ``co-occurrence'' or ``occurrence together''
in your study. In our examples you consider questions posed to one
and the same person (and responses obtained from one and the same
person) as co-occurring. You also know the rules, so you always use
different contexts, whatever your choice of the labels for them, for
the measurements that do not co-occur. Suppose I simplify your design
by forgetting about the presentation modes. I ask one of two questions,
$A$ or $B$, of randomly chosen people. The objects being measured
then are $A$ and $B$ again, and we know that the measurements of
these objects never co-occur, hence they are stochastically unrelated.
Therefore the questions themselves (or any two labels corresponding
to them one-to-one) are the contexts here,
\[
R_{A}^{A},R_{B}^{B}.
\]

\noindent \qq I think I now understand the notion of stochastic unrelatedness
and your notation. But couldn't we also say in all such cases that
the two random variables, e.g., $R_{A}^{A},R_{B}^{B}$, are stochastically
independent? 

\noindent \pp This would be a common way of thinking of this situation.
But it is incorrect. One could only say, with some caution, that they
can always be treated \emph{as if }they were independent. We will
get to this later, when we consider the notion of a coupling. You
can, however, appreciate the difference between the situation when
a randomly chosen person is being asked one of two questions, $A$
or $B$, and the situation when a randomly chosen person is being
asked both these questions, $A$ and $B$. The two responses in the
latter case are random variables
\[
\left(R_{A}^{\left(A,B\right)},R_{B}^{\left(A,B\right)}\right),
\]
jointly distributed: one can define and estimate the probability of
the joint event $\left[R_{A}^{\left(A,B\right)}=\textnormal{Yes and }R_{B}^{\left(A,B\right)}=\textnormal{Yes}\right]$.
Suppose we find out that this probability equals the product of the
probability of $\left[R_{A}^{\left(A,B\right)}=\textnormal{Yes}\right]$
and the probability of $\left[R_{B}^{\left(A,B\right)}=\textnormal{Yes}\right]$
taken separately. Then we say that the random variables $R_{A}^{\left(A,B\right)}$
and $R_{B}^{\left(A,B\right)}$ are stochastically independent. This
is fundamentally different from the situation with $R_{A}^{A}$ and
$R_{B}^{B}$, when the joint event $\left[R_{A}^{A}=\textnormal{Yes and }R_{B}^{B}=\textnormal{Yes}\right]$
is simply undefined.\footnote{\textsc{Aut:} For a detailed discussion, see \cite{DK2014LNCSQualified}
and \cite{Dzhafarov2015}.}

\noindent \qq I see the difference. I still have questions about
the objects and contexts, but I think I should allow you to continue
your presentation of CbD. The double-notation, I understand, is only
a departure point. 

\noindent \pp Yes, it is. And we will indeed continue to address
your misgivings as we proceed. By now we have established a certain
picture of a system of measurements: it consists of stochastically
unrelated to each other islands (or bunches) of jointly distributed
random variables. The main idea of CbD is that these isolated bunches
can be characterized by exploring \emph{all possible couplings thereof
(}or\emph{ all possible joint distributions imposable on them) under
well-chosen constraints}. 

\noindent \qq What is a \emph{coupling}? 

\noindent \pp A \index{coupling}coupling for stochastically unrelated random variables
$X,Y,\ldots,Z$ is a jointly distributed $\left(\widetilde{X},\widetilde{Y},\ldots,\widetilde{Z}\right)$
in which $\widetilde{X}$ has the same distribution as $X$, $\widetilde{Y}$
has the same distribution as $Y$, and so on.\footnote{\textsc{Aut:} See \cite{Thorisson2000}. A traditional definition
of a coupling does not require that the random variables being coupled
be stochastically unrelated, but in the present context it is the
only application.} For instance, in our example with $R_{A}^{A}$ and $R_{B}^{B}$,
let the distributions of these random variables be defined by (with
$\Pr$ standing for probability) 
\[
\Pr\left[R_{A}^{A}=\textnormal{Yes}\right]=0.4,\Pr\left[R_{B}^{B}=\textnormal{Yes}\right]=0.7.
\]
The measurements $R_{A}^{A}$ and $R_{B}^{B}$ are stochastically
unrelated, so $\Pr\left[R_{A}^{A}=\textnormal{Yes and }R_{B}^{B}=\textnormal{Yes}\right]$
is undefined. To couple them means to create a new pair of random
variables, $\widetilde{R}_{A}^{A}$ and $\widetilde{R}_{B}^{B}$,
that are distributional copies of, respectively, $R_{A}^{A}$ and
$R_{B}^{B}$ but are jointly distributed. That is, 
\[
\Pr\left[\widetilde{R}_{A}^{A}=\textnormal{Yes}\right]=0.4,\Pr\left[\widetilde{R}_{B}^{B}=\textnormal{Yes}\right]=0.7,
\]
but, unlike the ``originals'' $R_{A}^{A}$ and $R_{B}^{B}$, their
distributional copies have a well-defined joint probability, 
\[
\Pr\left[\widetilde{R}_{A}^{A}=\textnormal{Yes and }\widetilde{R}_{B}^{B}=\textnormal{Yes}\right].
\]
It can be shown that this probability can have any value from 0.1
(maximally negative relation) to 0.4 (maximally positive relation).
The independent coupling, with this probability equal to $0.4\times0.7$,
is within this range. 

\noindent \qq And any of these values will define a pair $\left(\widetilde{R}_{A}^{A},\widetilde{R}_{B}^{B}\right)$
that is a coupling for $
{R}_{A}^{A}$ and ${R}_{B}^{B}$?

\noindent \pp Yes. The number of all possible couplings for a given
pair of random variables is typically infinite. For binary $X,Y$
(say, Yes/No ones) the only exceptions are the pairs with $\Pr\left[X=\textnormal{Yes}\right]$
or $\Pr\left[Y=\textnormal{Yes}\right]$ having the values 0 or 1.
For such a pair only one coupling is possible.

\noindent \qq Let me try an example with continuous distributions.
Take random variables $R$ and $S$ that are standard normally distributed.
Then any bivariate normally distributed $\left(\widetilde{R},\widetilde{S}\right)$
with standard normal marginals is a coupling of $R$ and $S$.

\noindent \pp Yes, and there are other couplings for these $R$ and
$S$ too: it can be any $\left(\widetilde{R},\widetilde{S}\right)$
whose joint distribution is well-defined, and whose individual (marginal)
distributions are standard normal.

\noindent \qq Just so that we don't focus on pairs of random variables
exclusively, what would be a coupling for the four random variables
$R_{A}^{V},R_{A}^{W},R_{B}^{V},R_{B}^{W}$ in our original example?
There we had two pairs (you called them ``bunches'') with jointly
distributed components, $\left(R_{A}^{V},R_{B}^{V}\right)$ and $\left(R_{A}^{W},R_{B}^{W}\right)$.

\noindent \pp Because of the latter, this example too can be presented
as involving just pairs: $\left(R_{A}^{V},R_{B}^{V}\right)$ is a
random variable too, unless we use the term very narrowly. But this
is not critical: it is the same thing to seek a coupling for $\left(R_{A}^{V},R_{B}^{V}\right)$
and $\left(R_{A}^{W},R_{B}^{W}\right)$, each with a known distribution,
and to seek a coupling for $R_{A}^{V},R_{A}^{W},R_{B}^{V},R_{B}^{W}$
in which you know the distributions of $\left(R_{A}^{V},R_{B}^{V}\right)$
and $\left(R_{A}^{W},R_{B}^{W}\right)$. The joint distribution of
$\left(R_{A}^{V},R_{B}^{V}\right)$ is determined by four probabilities
\[
\begin{array}{rcccc}
\left(R_{A}^{V},R_{B}^{V}\right): & \left(\textnormal{Yes},\textnormal{Yes}\right) & \left(\textnormal{Yes},\textnormal{No}\right) & \left(\textnormal{No},\textnormal{Yes}\right) & \left(\textnormal{No},\textnormal{No}\right)\\
\textnormal{probability}: & p_{YY} & p_{YN} & p_{NY} & p_{NN}
\end{array},
\]
with the probabilities summing to 1, of course. And the joint distribution
of $\left(R_{A}^{W},R_{B}^{W}\right)$ is determined analogously,
\[
\begin{array}{rcccc}
\left(R_{A}^{W},R_{B}^{W}\right): & \left(\textnormal{Yes},\textnormal{Yes}\right) & \left(\textnormal{Yes},\textnormal{No}\right) & \left(\textnormal{No},\textnormal{Yes}\right) & \left(\textnormal{No},\textnormal{No}\right)\\
\textnormal{probability}: & q_{YY} & q_{YN} & q_{NY} & q_{NN}
\end{array}.
\]
To couple these pairs (equivalently, to couple all four random variables
$R_{A}^{V},R_{A}^{W},R_{B}^{V},R_{B}^{W}$) means to create a quadruple
$\left(\widetilde{R}_{A}^{V},\widetilde{R}_{A}^{W},\widetilde{R}_{B}^{V},\widetilde{R}_{B}^{W}\right)$
with jointly distributed components such that the distributions of
the pairs $\left(\widetilde{R}_{A}^{V},\widetilde{R}_{B}^{V}\right)$
and $\left(\widetilde{R}_{A}^{W},\widetilde{R}_{B}^{W}\right)$ are
the same as those of $\left(R_{A}^{V},R_{B}^{V}\right)$ and $\left(R_{A}^{W},R_{B}^{W}\right)$,
respectively. For instance,
$$\Pr\left[\widetilde{R}_{A}^{V}=\textnormal{Yes and }\widetilde{R}_{B}^{V}=\textnormal{No}\right]=p_{YN},\\
  \Pr\left[\widetilde{R}_{A}^{W}=\textnormal{No and }\widetilde{R}_{B}^{W}=\textnormal{No}\right]=q_{NN},\textnormal{ etc.}$$

\noindent \qq And to create a quadruple $\left(\widetilde{R}_{A}^{V},\widetilde{R}_{A}^{W},\widetilde{R}_{B}^{V},\widetilde{R}_{B}^{W}\right)$
with jointly distributed components means ...

\noindent \pp It means to assign to each of the 16 possible quadruples
of values a probability value:
\begin{small}
\[
\begin{array}{rcccc}
(\widetilde{R}_{A}^{V},\widetilde{R}_{A}^{W},\widetilde{R}_{B}^{V},\widetilde{R}_{B}^{W}): & \left(\textnormal{Yes},\textnormal{Yes},\textnormal{Yes},\textnormal{Yes}\right) & \left(\textnormal{Yes},\textnormal{Yes},\textnormal{Yes},\textnormal{No}\right) & \ldots & \left(\textnormal{No},\textnormal{No},\textnormal{No},\textnormal{No}\right)\\
\textnormal{probability}: & r_{YYYY} & r_{YYYN} & \ldots & r_{NNNN}
\end{array}.
\]
To form a coupling, these probabilities should agree with the observed
$p$ and $q$ values: e.g., 
\[
\sum_{i,j\in\left\{ \textnormal{Yes},\textnormal{No}\right\} }r_{YiNj}=p_{YN},\;\sum_{i,j\in\left\{ \textnormal{Yes},\textnormal{No}\right\} }r_{iNjN}=q_{NN}.
\]
\end{small}

\noindent \qq And, of course, this can generally be done in an infinite
number of ways. I think it is clear. You said earlier that you wanted
to characterize the isolated bunches of random variables by exploring
\emph{all possible couplings of these bunches under well-chosen constraints}.
Tell me now what you mean by the ``well-chosen constraints'' for
couplings.

\noindent \pp The ``well-chosen constraints'' depend on (and determine)
the aspect of the system of unrelated to each other bunches you want
to characterize. If we are interested in contextuality, the idea (arguably,
the most original idea in the CbD approach) is to look for \emph{couplings
in which the measurements of one and the same object under different
conditions are equal to each other with as high a probability as possible}.
Contextuality is determined by computing these highest probabilities
for each object in isolation and then determining if they are compatible
with the observed bunches of measurements. 

\noindent \qq I assume you want to elaborate.

\noindent \pp I will. But I think we will relegate this to our next
conversation.

\section*{Conversation 2}

\noindent \pp My dear Interlocutor, you asked me to elaborate what
I said about the measurements of an object in different contexts being
equal to each other with as high a probability as possible. First
of all, let me emphasize that the probability of being equal to each
other applies to the \index{coupling}\emph{couplings} rather than the coupled random
variables themselves. When I find a bivariate-normally distributed
coupling $\left(\widetilde{R},\widetilde{S}\right)$ for standard
normally distributed $R$ and $S$, I do not make $R$ and $S$ jointly
distributed, I merely create jointly distributed ``copies'' of $R$
and $S$. And there are generally an infinite set of such couplings.
Among them there is one coupling, with the correlation between $\widetilde{R}$
and $\widetilde{S}$ equal to 1 (defining a degenerate bivariate-normal
distribution), in which $\Pr\left[\widetilde{R}=\widetilde{S}\right]$
has the highest possible value (in this case, 1). It is called a \index{coupling!---, maximal}\emph{maximal
coupling}. If $R$ and $S$ are normally distributed but with different
means and/or variances, then the maximal couplings still exist, but
not among bivariate normally distributed $\left(\widetilde{R},\widetilde{S}\right)$,
and the highest possible probability for $\left[\widetilde{R}=\widetilde{S}\right]$
is less than 1.

\noindent \qq Let me switch back to our original example to understand
this. In $R_{A}^{V},R_{A}^{W},R_{B}^{V},R_{B}^{W}$ we have $R_{A}^{V}$
and $R_{A}^{W}$ measuring the same object $A$ in two contexts. We
take them and look for a coupling $\left(\widehat{R}_{A}^{V},\widehat{R}_{A}^{W}\right)$
for them, forgetting for the time being all about the remaining two
variables. We ask the question: what is the maximal possible probability
for the event $\left[\widehat{R}_{A}^{V}=\widehat{R}_{A}^{W}\right]$?
We know that, by definition of a coupling,
\[
\begin{array}{c}
\begin{array}{c}
\Pr\left[\widehat{R}_{A}^{V}=1\right]=\Pr\left[R_{A}^{V}=1\right]=p_{A}^{V},\\
\\
\Pr\left[\widehat{R}_{A}^{W}=1\right]=\Pr\left[R_{A}^{W}=1\right]=p_{A}^{W}.
\end{array}\end{array}
\]
I assume this allows me to compute the maximal possible value for
$\Pr\left[\widehat{R}_{A}^{V}=\widehat{R}_{A}^{W}\right]$?

\noindent \pp Yes. This maximal value is $1-\left|p_{A}^{V}-p_{A}^{W}\right|$.
It is very easy to prove.\footnote{\textsc{Aut:} See, e.g., \cite{DKL2015FooP}.}

\noindent \qq I will take you word for it. So we have
\[
\max_{\substack{\textnormal{all couplings}\\
\left(\widehat{R}_{A}^{V},\widehat{R}_{A}^{W}\right)
}
}\Pr\left[\widehat{R}_{A}^{V}=\widehat{R}_{A}^{W}\right]=1-\left|p_{A}^{V}-p_{A}^{W}\right|.
\]
After that I forget all about $R_{A}^{V}$ and $R_{A}^{W}$ and focus
on $R_{B}^{V}$ and $R_{B}^{W}$, the other two measurements of one
and the same object in two contexts. By analogy, 
\[
\max_{\substack{\textnormal{all couplings}\\
\left(\widehat{R}_{B}^{V},\widehat{R}_{B}^{W}\right)
}
}\Pr\left[\widehat{R}_{B}^{V}=\widehat{R}_{B}^{W}\right]=1-\left|p_{B}^{V}-p_{B}^{W}\right|.
\]
How does one proceed from here?

\noindent \pp Now we have to take all four our random variables and
construct a coupling $\left(\widetilde{R}_{A}^{V},\widetilde{R}_{A}^{W},\widetilde{R}_{B}^{V},\widetilde{R}_{B}^{W}\right)$
for them. We have already discussed how we do this. Except in special
cases, there is an infinity of such couplings. What we are now interested
in is whether among all these couplings there is at least one in which
\[
\begin{array}{c}
\Pr\left[\widetilde{R}_{A}^{V}=\widetilde{R}_{A}^{W}\right]=1-\left|p_{A}^{V}-p_{A}^{W}\right|\\
\textnormal{and}\\
\Pr\left[\widetilde{R}_{B}^{V}=\widetilde{R}_{B}^{W}\right]=1-\left|p_{B}^{V}-p_{B}^{W}\right|.
\end{array}
\]
If the answer to this question is affirmative, then we say that the
system of measurements, in this case comprised of $\left(R_{A}^{V},R_{B}^{V}\right)$
and $\left(R_{A}^{W},R_{B}^{W}\right)$, is \index{system of measurements!---, (non)contextual}\emph{noncontextual}.
If there is no such couplings, then the system is \index{system of measurements!---,  (non)contextual}\emph{contextual}.

\noindent \qq Let me first see how this works if 
\[
p_{A}^{V}=p_{A}^{W}\textnormal{ and }p_{B}^{V}=p_{B}^{W}.
\]
 The maximal probabilities of $\left[\widehat{R}_{A}^{V}=\widehat{R}_{A}^{W}\right]$
and of $\left[\widehat{R}_{B}^{V}=\widehat{R}_{B}^{W}\right]$ then
are equal to 1, for both $A$ and $B$. So, if I can find a coupling
$\left(\widetilde{R}_{A}^{V},\widetilde{R}_{A}^{W},\widetilde{R}_{B}^{V},\widetilde{R}_{B}^{W}\right)$
in which $\widetilde{R}_{A}^{V}$ is always the same as $\widetilde{R}_{A}^{W}$
and $\widetilde{R}_{B}^{V}$ is always the same as $\widetilde{R}_{B}^{W}$,
then the system is noncontextual. 

\noindent \pp This is in fact the traditional understanding of contextuality
(expressed in the language of CbD).\footnote{\textsc{Aut:} See \cite{DK2014Scripta,DK2014LNCSQualified}.} If $\widetilde{R}_{A}^{V}$ and $\widetilde{R}_{A}^{W}$ are always
the same, one can say that the measurement of $A$ does not depend
on what the context is, $V$ or $W$; and analogously for $\widetilde{R}_{B}^{V}$
and $\widetilde{R}_{B}^{W}$. 

\noindent \qq The adjective ``noncontextual'' here seems intuitive
to me. Let me now consider the case when $p_{A}^{V}\not=p_{A}^{W}$,
i.e., the maximal probability $1-\left|p_{A}^{V}-p_{A}^{W}\right|<1$.
So we begin by computing ... But wait: the measurements of $A$ here
have different distributions in context $V$ and in context $W$,
so these measurement are context-dependent. Why don't we declare this
system of measurements contextual, without computing anything else?

\noindent \pp You are touching on a subtle conceptual and terminological
issue. Nothing prevents one from calling this system contextual, but
in the terminology of CbD it is called \index{system of measurements!---, (in)consistently connected}\emph{inconsistently connected}
(and if $p_{A}^{V}=p_{A}^{W}$ and $p_{B}^{V}=p_{B}^{W}$, the system
is \index{system of measurements!---, (in)consistently connected}\emph{consistently connected}). If you insist on using the word
``contextuality'' whenever a system has $p_{A}^{V}\not=p_{A}^{W}$
or $p_{B}^{V}\not=p_{B}^{W}$, call this ``contextuality-1.'' Or
use another qualifier, but distinguish this form of contextuality
from the contextuality in the sense of CbD (call it ``contextuality-2''
if you like). It is the form of contextuality that may exist on top
of the ``contextuality-1.'' 

\noindent \qq I still have misgivings, but we can return to this
later. Let me resume my attempt to understand how this ``contextuality-2''
works in the case $p_{A}^{V}\not=p_{A}^{W}$ and/or $p_{B}^{V}\not=p_{B}^{W}$.
We compute the maximal probability of $\left[\widehat{R}_{A}^{V}=\widehat{R}_{A}^{W}\right]$
across all couplings $\left(\widehat{R}_{A}^{V},\widehat{R}_{A}^{W}\right)$
for $R_{A}^{V}$ and $R_{A}^{W}$; and, separately, we compute the
maximal probability of $\left[\widehat{R}_{B}^{V}=\widehat{R}_{B}^{W}\right]$
across all couplings $\left(\widehat{R}_{B}^{V},\widehat{R}_{B}^{W}\right)$
for $R_{B}^{V}$ and $R_{B}^{W}$. These probabilities, you tell me,
are $1-\left|p_{A}^{V}-p_{A}^{W}\right|$ and $1-\left|p_{B}^{V}-p_{B}^{W}\right|$,
respectively. Then we look at all possible couplings $\left(\widetilde{R}_{A}^{V},\widetilde{R}_{A}^{W},\widetilde{R}_{B}^{V},\widetilde{R}_{B}^{W}\right)$
for all four of our random variables, and for each of them we compute
the probabilities of $\left[\widetilde{R}_{A}^{V}=\widetilde{R}_{B}^{V}\right]$
and of $\left[\widetilde{R}_{A}^{W}=\widetilde{R}_{B}^{W}\right]$. 

\noindent \pp It is clear that these probabilities cannot exceed
the values $1-\left|p_{A}^{V}-p_{A}^{W}\right|$ and $1-\left|p_{B}^{V}-p_{B}^{W}\right|$,
respectively --- because every sub-coupling $\left(\widetilde{R}_{A}^{V},\widetilde{R}_{A}^{W}\right)$
of $\left(\widetilde{R}_{A}^{V},\widetilde{R}_{A}^{W},\widetilde{R}_{B}^{V},\widetilde{R}_{B}^{W}\right)$
is also one of the possible couplings $\left(\widehat{R}_{A}^{V},\widehat{R}_{A}^{W}\right)$
for $R_{A}^{V}$ and $R_{A}^{W}$ taken separately; and analogously
for $\left(\widetilde{R}_{B}^{V},\widetilde{R}_{B}^{W}\right)$ and
$\left(\widehat{R}_{B}^{V},\widehat{R}_{B}^{W}\right)$.

\noindent \qq Yes, I see this. Now, if in some of the couplings $\left(\widetilde{R}_{A}^{V},\widetilde{R}_{A}^{W},\widetilde{R}_{B}^{V},\widetilde{R}_{B}^{W}\right)$
both these probabilities are achieved, then we say the system is noncontextual
(lacks ``contextuality-2''). Otherwise it is contextual.

\noindent \pp Yes. The intuition here is that there is something
in the relationship between $\left(R_{A}^{V},R_{B}^{V}\right)$ and
$\left(R_{A}^{W},R_{B}^{W}\right)$ that cannot be reduced to the
separate effects of the context changes on the responses to $A$ and
on the responses to $B$. 

\noindent \qq Wait, wait. Couldn't we then simply reformulate the
problem by taking $\left(A,B\right)$ as a single object? Then we
would have two stochastically unrelated random variables
\[
R_{\left(A,B\right)}^{V},R_{\left(A,B\right)}^{W},
\]
each with four possible values, (Yes-Yes), (Yes-No), etc. We will
have contextuality if their distributions are different. This would
be ``contextuality-1,'' of course.

\noindent \pp This is definitely a possible approach. I have mentioned
already that objects and contexts are primitives of the CbD theory,
which means that the theory does not dictate their choice. The only
constraint imposed by the theory is that the random variables measured
in the same context have a joint distribution, while random variables
in different contexts do not. If you choose a single object in two
contexts, as you have proposed, you will simply be dealing with a
different problem. The system comprised of $R_{\left(A,B\right)}^{V}$
and $R_{\left(A,B\right)}^{W}$ may or may not be inconsistently connected
(I suggest we stick to this term instead of ``contextuality-1''
and the like), but irrespective of this, it is noncontextual. If it
is inconsistently connected, then the system we considered previously,
$\left(R_{A}^{V},R_{B}^{V}\right)$ and $\left(R_{A}^{W},R_{B}^{W}\right)$,
may or may not be consistently connected, and in either case it can
be contextual or noncontextual. 

\noindent \qq Then I was wrong: inconsistent connectedness in $\left\{ R_{\left(A,B\right)}^{V},R_{\left(A,B\right)}^{W}\right\}$
does not predict or account for the contextuality in $\left\{ \left(R_{A}^{V},R_{B}^{V}\right),\left(R_{A}^{W},R_{B}^{W}\right)\right\} $. 

\noindent \pp No, it does not. The system $\left\{ \left(R_{A}^{V},R_{B}^{V}\right),\left(R_{A}^{W},R_{B}^{W}\right)\right\} $
can be shown to be noncontextual if and only if
\[
\left|\left\langle R_{A}^{V}R_{B}^{V}\right\rangle -\left\langle R_{A}^{W}R_{B}^{W}\right\rangle \right|\leq\left|\left\langle R_{A}^{V}\right\rangle -\left\langle R_{A}^{W}\right\rangle \right|+\left|\left\langle R_{B}^{V}\right\rangle -\left\langle R_{B}^{W}\right\rangle \right|,
\]
where $\left\langle \cdot\right\rangle $ is expected value.\footnote{\textsc{Aut:} This is an example of a \index{system of measurements!---, cyclic}cyclic system with $n=2$,
as defined in Conversation 3.} You can easily verify that this inequality may hold or fail with
the distributions of $R_{\left(A,B\right)}^{V}$ and $R_{\left(A,B\right)}^{W}$
being different.

\noindent \qq I wonder: even if we get a completely different system
by doing this, is it always possible to get rid of contextuality in
the sense of CbD by redefining the objects?

\noindent \pp The answer to this is yes, but generally not by grouping
the objects together, as in $R_{\left(A,B\right)}^{V},R_{\left(A,B\right)}^{W}$.
Consider, e.g., a system involving three objects $q,q',q''$ and three
contexts $c,c',c''$, combined in the measurements as follows:\footnote{\textsc{Aut:} This is an example of a \index{system of measurements!---, cyclic}cyclic system with $n=3$.}
\[
\left(R_{q}^{c},R_{q'}^{c}\right),\left(R_{q'}^{c'},R_{q''}^{c'}\right),\left(R_{q''}^{c''},R_{q}^{c''}\right).
\]
As always, the pairs of random variables labeled by the same context
are jointly distributed, and different pairs are stochastically unrelated.
One cannot now put all three objects together as a single object.
Instead one can do something universally applicable: taking a measurement's
context as part of the identity of the measurement's object. In our
case this means replacing $q$ in $R_{q}^{c}$ with $\left(q,c\right)$,
replacing $q$ in $R_{q}^{c''}$ with $\left(q,c''\right)$, etc.
We will get then in place of the system above a new system
\[
\left(R_{\left(q,c\right)}^{c},R_{\left(q',c\right)}^{c}\right),\left(R_{\left(q',c'\right)}^{c'},R_{\left(q'',c'\right)}^{c'}\right),\left(R_{\left(q'',c''\right)}^{c''},R_{\left(q,c''\right)}^{c''}\right).
\]
In this system no two measurements share their object, and the
system is readily seen as trivially noncontextual (and also consistently
connected, again in the trivial sense). 

\noindent \qq This universal trick then consists in declaring any
\index{measurement!--- object}object in a new \index{measurement!--- context}context to be a new object. For instance, a question
presented verbally and the same (in content) question presented in
writing are different questions. 

\noindent \pp Yes, and there is nothing incorrect about this, at
least not from the point of view of CbD. It just makes the issue of
contextuality uninteresting. 

\noindent \qq So we will not use this trick for the sake of keeping
our discussion interesting. However, philosophically speaking, it
may very well be the case that finding a system of measurements contextual
means that ``the same'' objects in different contexts are not really
the same. 

\noindent \pp Perhaps. But I find such philosophical formulations
unsatisfactory. We need a language rich enough to lead to interesting
classifications and quantifications of contextuality. The language
of CbD is rich enough. Trivial renaming of all objects into object-contexts
is not.

\noindent \qq I agree. I think I understand the definition of contextuality
in CbD. However, I may need more persuasion to accept it. Let me return
to my misgivings about ``contextuality-1'' and ``contextuality-2.''
Why do we need the latter?

\noindent \pp Let me remind to you that ``contextuality-1'' is
inconsistent connectedness: for some objects, their measurements have
different distributions in different contexts. We may very well have
a consistently connected system, however, without ``contextuality-1.''
Will we simply declare it noncontextual since all contextuality is
``contextuality-1''? Again, one can say this if one so wishes, but
this terminology will not change the fact that there is an important
distinction within the class of consistently connected systems of
measurements. 

\noindent \qq Please remind me what this distinction is.

\noindent \pp If the distribution of the measurements of a given
object is the same across all contexts involving this object, then
it is possible to couple these measurements so that their copies in
the coupling are equal to each other with probability 1. Let's call
this the \index{coupling!---, identity}\emph{identity coupling}. Now, there are two possibilities.
There can be a consistently connected system that has a coupling in
which this probability 1 is achieved for all objects; in other words
the identity couplings for different objects can all be put together
so that they are compatible with the observed distributions of the
bunches of measurements in each of the contexts. And there can be
systems in which such couplings do not exist: the observed bunches
of measurements are not compatible with the identity couplings for
all the objects. This is an important distinction, and it is captured
by calling the systems of the latter kind contextual. In quantum physics
this distinction is related to such questions as the (non)existence
of hidden variables of which all observed random variables in an experiment
are functions.\footnote{\textsc{Aut:} This is, e.g., how the problem was formulated by \cite{Bell1964}.
The possibility of reformulating this in terms of the (non)existence
of certain couplings (without using this concept explicitly) was realized
later, in \cite{SuppesZanotti1981} and \cite{Fine1982b}.} 

\noindent \qq Yes, I agree this distinction is important. But it
seems to me in quantum physics the case of consistent connectedness
is the main if not the only case to consider.\footnote{\textsc{Aut:} It may be a prevalent case but definitely not the only
one: see, e.g., \cite{Bacciagaluppi2015}.} If we retain the traditional definition of contextuality for such
systems (perhaps in the CbD formulation), do we need to extend it
to inconsistent connectedness? You said the contextuality in the sense
of CbD exists ``on top of'' inconsistent connectedness. Couldn't
we, however, simply ignore it? In other words, couldn't we have a
single notion of contextuality, which coincides with ``contextuality-2''
for consistently connected systems and with ``contextuality-1''
otherwise? I have almost asked this question before, but we digressed
(or at least I see it now as a digression) into the discussion of
how one can define objects and contexts. 

\noindent \pp Let me think of how to respond to this, and we will
return to this in our next conversation.

\section*{Conversation 3}

\noindent \pp My dear Interlocutor, to defend a definition is a difficult
task. A good definition of a term should be intuitively plausible
(although sometimes one's intuition itself should be ``educated''
to make it plausible), it should include as special cases all examples
and situations that are traditionally considered to fall within the
scope of the term, it should lead to productive development (to allow
one to prove nontrivial theorems), and have a growing set of applications.
I believe contextuality in the sense of CbD satisfies all these desiderata,
but I may be unable to discuss them with you comprehensively.

\noindent \qq Let us try intuitive plausibility.

\noindent \pp One argument I find persuasive is appealing to ``small''
\index{system of measurements!---, (in)consistently connected}inconsistencies added to \index{system of measurements!---, (in)consistently connected}consistently connected systems with ``large''
contextuality. Contextuality in CbD can be rigorously quantified,\footnote{\textsc{Aut:} See \cite{DKL2015FooP}; \cite{KujalaDzhafarovLarsson2015};
\cite{KujalaDzhafarov2015}; \cite{deBarrosDzhafarovKujalaOas2015}.} but I will only need intuitive guidance to present the argument.
Consider our system $\left\{ \left(R_{A}^{V},R_{B}^{V}\right),\left(R_{A}^{W},R_{B}^{W}\right)\right\} $.
You may recall that the criterion (necessary and sufficient condition)
for noncontextuality here is given by the inequality
\[
\left|\left\langle R_{A}^{V}R_{B}^{V}\right\rangle -\left\langle R_{A}^{W}R_{B}^{W}\right\rangle \right|\leq\left|\left\langle R_{A}^{V}\right\rangle -\left\langle R_{A}^{W}\right\rangle \right|+\left|\left\langle R_{B}^{V}\right\rangle -\left\langle R_{B}^{W}\right\rangle \right|.
\]
Your proposal is to accept this formula only for consistently connected
systems, when
\[
\left|\left\langle R_{A}^{V}\right\rangle -\left\langle R_{A}^{W}\right\rangle \right|+\left|\left\langle R_{B}^{V}\right\rangle -\left\langle R_{B}^{W}\right\rangle \right|=0.
\]
In this case the system is contextual if and only if 
\[
\left|\left\langle R_{A}^{V}R_{B}^{V}\right\rangle -\left\langle R_{A}^{W}R_{B}^{W}\right\rangle \right|>0.
\]
Now, the largest possible value of the last expression is 2, and,
I think you would agree, it is reasonable to say that the system $\left\{ \left(R_{A}^{V},R_{B}^{V}\right),\left(R_{A}^{W},R_{B}^{W}\right)\right\} $
with this value equal to 2 exhibits the greatest possible degree of
contextuality, given the consistent connectedness. 

\noindent \qq Is this maximum value 2 compatible with consistent
connectedness?

\noindent \pp Yes, it is. See these two distributions:

\begin{center}
\begin{tabular}{|c|c|c|c}
\multicolumn{1}{c}{} & \multicolumn{1}{c}{} & \multicolumn{1}{c}{} & \tabularnewline
\cline{2-3} 
\multicolumn{1}{c|}{$\begin{array}{c}
\\
\\
\end{array}$} & \!$R_{B}^{V}=+1$\! & \!$R_{B}^{V}=-1$\! & \tabularnewline
\hline 
\!$R_{A}^{V}=+1\!\!\begin{array}{c}
\\
\\
\end{array}$ & $\frac{1}{2}$ & $0$ & \multicolumn{1}{c|}{$\frac{1}{2}$}\tabularnewline
\hline 
\!$R_{A}^{V}=-1\!\!\begin{array}{c}
\\
\\
\end{array}$ & $0$ & $\frac{1}{2}$ & \multicolumn{1}{c|}{$\frac{1}{2}$}\tabularnewline
\hline 
\multicolumn{1}{c|}{$\begin{array}{c}
\\
\\
\end{array}$} & $\frac{1}{2}$ & $\frac{1}{2}$ & \tabularnewline
\cline{2-3} 
\multicolumn{1}{c}{} & \multicolumn{1}{c}{} & \multicolumn{1}{c}{} & \tabularnewline
\end{tabular}$\quad$%
\begin{tabular}{|c|c|c|c}
\multicolumn{1}{c}{} & \multicolumn{1}{c}{} & \multicolumn{1}{c}{} & \tabularnewline
\cline{2-3} 
\multicolumn{1}{c|}{$\begin{array}{c}
\\
\\
\end{array}$} & \!$R_{B}^{W}=+1$\! & \!$R_{B}^{W}=-1$\! & \tabularnewline
\hline 
\!$R_{A}^{W}=+1\!\!\begin{array}{c}
\\
\\
\end{array}$ & $0$ & $\frac{1}{2}$ & \multicolumn{1}{c|}{$\frac{1}{2}$}\tabularnewline
\hline 
\!$R_{A}^{W}=-1\!\!\begin{array}{c}
\\
\\
\end{array}$ & $\frac{1}{2}$ & $0$ & \multicolumn{1}{c|}{$\frac{1}{2}$}\tabularnewline
\hline 
\multicolumn{1}{c|}{$\begin{array}{c}
\\
\\
\end{array}$} & $\frac{1}{2}$ & $\frac{1}{2}$ & \tabularnewline
\cline{2-3} 
\multicolumn{1}{c}{} & \multicolumn{1}{c}{} & \multicolumn{1}{c}{} & \tabularnewline
\end{tabular}
\par\end{center}

\ppp The computations yield all the expected values $\left\langle R_{A}^{V}\right\rangle ,\left\langle R_{A}^{W}\right\rangle ,\left\langle R_{B}^{V}\right\rangle ,\left\langle R_{B}^{W}\right\rangle $
equal to zero, $\left\langle R_{A}^{V}R_{B}^{V}\right\rangle =1$
and $\left\langle R_{A}^{W}R_{B}^{W}\right\rangle =-1$.

\noindent \qq I see. Please continue with your example.

\noindent \pp Let us now introduce a minuscule inconsistency, say, 

\begin{center}
\begin{tabular}{|c|c|c|c}
\multicolumn{1}{c}{} & \multicolumn{1}{c}{} & \multicolumn{1}{c}{} & \tabularnewline
\cline{2-3} 
\multicolumn{1}{c|}{$\begin{array}{c}
\\
\\
\end{array}$} & \!\!$R_{B}^{V}=+1$\!\! & \!\!$R_{B}^{V}=-1$\!\! & \tabularnewline
\hline 
\!\!$R_{A}^{V}=+1\!\!\!\begin{array}{c}
\\
\\
\end{array}$ & $\frac{1}{2}$ & $\varepsilon$ & \multicolumn{1}{c|}{\!\!$\frac{1}{2}+\varepsilon$\!\!}\tabularnewline
\hline 
\!\!$R_{A}^{V}=-1\!\!\!\begin{array}{c}
\\
\\
\end{array}$ & $0$ & $\frac{1}{2}-\varepsilon$ & \multicolumn{1}{c|}{\!\!$\frac{1}{2}-\varepsilon$\!\!}\tabularnewline
\hline 
\multicolumn{1}{c|}{$\begin{array}{c}
\\
\\
\end{array}$} & $\frac{1}{2}$ & $\frac{1}{2}$ & \tabularnewline
\cline{2-3} 
\multicolumn{1}{c}{} & \multicolumn{1}{c}{} & \multicolumn{1}{c}{} & \tabularnewline
\end{tabular}$\quad$%
\begin{tabular}{|c|c|c|c}
\multicolumn{1}{c}{} & \multicolumn{1}{c}{} & \multicolumn{1}{c}{} & \tabularnewline
\cline{2-3} 
\multicolumn{1}{c|}{$\begin{array}{c}
\\
\\
\end{array}$} & \!\!$R_{B}^{W}=+1$\!\! & \!\!$R_{B}^{W}=-1$\!\! & \tabularnewline
\hline 
\!\!$R_{A}^{W}=+1\!\!\!\begin{array}{c}
\\
\\
\end{array}$ & $0$ & $\frac{1}{2}$ & \multicolumn{1}{c|}{$\frac{1}{2}$}\tabularnewline
\hline 
\!\!$R_{A}^{W}=-1\!\!\!\begin{array}{c}
\\
\\
\end{array}$ & $\frac{1}{2}$ & $0$ & \multicolumn{1}{c|}{$\frac{1}{2}$}\tabularnewline
\hline 
\multicolumn{1}{c|}{$\begin{array}{c}
\\
\\
\end{array}$} & $\frac{1}{2}$ & $\frac{1}{2}$ & \tabularnewline
\cline{2-3} 
\multicolumn{1}{c}{} & \multicolumn{1}{c}{} & \multicolumn{1}{c}{} & \tabularnewline
\end{tabular}
\par\end{center}

\ppp The only difference of this system from the previous one is
that now $\left\langle R_{A}^{V}\right\rangle =2\varepsilon$ rather
than 0 and $\left\langle R_{A}^{V}R_{B}^{V}\right\rangle =1-2\varepsilon$
rather than 1; but $\varepsilon$ can be chosen arbitrarily small.
The system therefore can be made arbitrarily close to the previous
one. If I continue to follow your proposal, however, I should abandon
the joint expectations altogether and focus on the marginals only:
the system is contextual now simply because it is inconsistently connected,
\[
\left|\left\langle R_{A}^{V}\right\rangle -\left\langle R_{A}^{W}\right\rangle \right|+\left|\left\langle R_{B}^{V}\right\rangle -\left\langle R_{B}^{W}\right\rangle \right|=2\varepsilon>0.
\]
But you would probably agree that if $\varepsilon$ is minuscule,
the degree of contextuality in the system is minuscule too, wouldn't
you? 

\noindent \qq Indeed I would. And I can guess the rest of your argument
too. When $\varepsilon$ is very small but nonzero, the system has
a small degree of contextuality (which will be ``contextuality-1'').
As you make $\varepsilon$ smaller and smaller, the contextuality
gets smaller and smaller. But as soon as $\varepsilon$ reaches zero,
the contextuality jumps from the limiting zero value to the maximal
possible value (because now it is ``contextuality-2''). It is a
strange behavior, I should admit.

\noindent \pp Precisely. I conclude that your concept of contextuality
is not well-formed. If we distinguish inconsistent connectedness from
contextuality in accordance with CbD, however, the problem disappears.
The degree of contextuality in the second system, if $\varepsilon$
is very small, is only slightly smaller than the degree of contextuality
in the first system. The inequality
\[
2=\left|\left\langle R_{A}^{V}R_{B}^{V}\right\rangle -\left\langle R_{A}^{W}R_{B}^{W}\right\rangle \right|>\left|\left\langle R_{A}^{V}\right\rangle -\left\langle R_{A}^{W}\right\rangle \right|+\left|\left\langle R_{B}^{V}\right\rangle -\left\langle R_{B}^{W}\right\rangle \right|=0
\]
changes into
\[
2-2\varepsilon=\left|\left\langle R_{A}^{V}R_{B}^{V}\right\rangle -\left\langle R_{A}^{W}R_{B}^{W}\right\rangle \right|>\left|\left\langle R_{A}^{V}\right\rangle -\left\langle R_{A}^{W}\right\rangle \right|+\left|\left\langle R_{B}^{V}\right\rangle -\left\langle R_{B}^{W}\right\rangle \right|=2\varepsilon.
\]

\noindent \qq I agree this feature speaks in favor of the CbD concept.
Does this mean, however, that inconsistent connectedness and contextuality
(or ``contextuality-1'' and ``contextuality-2,'' even if you don't
like this terminology) have fundamentally different ontologies?

\noindent \pp It is a question to which I do not have a definitive
answer. Inconsistent connectedness in most, if not all cases have
trivial and well-understood causes: conditions under which measurements
are made affect these measurements, either through physical interference
or through context-dependent measurement biases. In the example with
written and verbal questions, reading a question invokes very different
psychological processes than hearing it asked: there is nothing remarkable
in the distributions of responses in the two cases being different.
Or consider a formally identical but empirically different example:
replace the presentation mode with order in which the two questions
are asked, so that instead of $V$ we have context $A\rightarrow B$
and instead of $W$ the context $B\rightarrow A$. In this case it
is natural to expect that the first question affects a person's response
to the second question.\footnote{\textsc{Aut:} See \cite{Moore2002}; \cite{Wang2014}; \cite{DZK2015}.}
In physics, it is common that different contexts correspond to different
experimental set-ups, so that one and the same object in different
contexts is simply measured differently.\footnote{
{\textsc{Aut:}} See the discussion of an experiment by \cite{Lapkiewicz2011}
in \cite{KujalaDzhafarovLarsson2015}.
}

\noindent \qq And you claim that the causes for contextuality are
different?

\noindent \pp At least they may be different, and they are definitely
different in some cases. Take the famous Alice-Bob experiment (more
formally known as the EPR/Bohm or EPR/Bell paradigm), in which Alice
measures spins in a particle 1 and Bob measures spins in a particle
2. The two particles are \emph{entangled}, meaning that they were
created in a special, singular state that makes Alice's and Bob's
spins that are measured in the opposite directions perfectly correlated.
Alice and Bob make their measurements simultaneously from some Charlie's
point of view, and this precludes any information traveling from Alice
to Bob or vice versa. Nevertheless, we have a clear case of contextuality
(``contextuality-2'') in this case.\footnote{\textsc{Aut:} See, e.g., \cite{DK2014FOOP}.}
If we modify this experiment so that Alice and Bob (from Charlie's
point of view) make their measurement with an interval between them
that allows for signaling, and if we assume that some form of signaling
is indeed effected, then we may have distributions of Alice's measurement
depending on Bob's settings and/or vice versa. This would be inconsistent
connectedness. But contextuality may still be measurable ``on top
of'' this inconsistency.\footnote{\textsc{Aut:} See the chapter by Kujala and Dzhafarov in this volume.}

\noindent \qq But you don't know if contextuality is always so different
in nature from inconsistent connectedness, do you?

\noindent \pp No, which is why I said I did not have a definitive
answer. 

\noindent \qq Do we know if contextuality of the kind we find in
quantum physics also exists in non-physical systems? Perhaps in human
behavior?

\noindent \pp No, we don't. A recent analysis of available experimental
data in psychology seems to suggest that the answer is negative.\footnote{\textsc{Aut:} See \cite{DZK2015}.}
But we can't know for sure, because in psychology we lack a theory
analogous to quantum mechanics and have to grope in darkness trying
now this and then that.

\noindent \qq So it is possible that contextuality only exists in
quantum physics? This would be disappointing, wouldn't it?

\noindent \pp Not necessarily. Lack of contextuality, if it can be
formulated as a general principle in some domain, allows us to predict
outcomes of experiments, or at least predict what outcomes are not
possible. In psychology this hypothetical principle would, in a sense,
create a general theory that we otherwise lack.

\noindent \qq I see. Well, we have covered a lot of ground in our
conversations. Would you like to summarize?

\noindent \pp I could summarize the definition of contextuality in
a more formal way. I recall you did not like informal presentations.

\noindent \qq Please do.

\noindent \pp The primitive concepts of the theory are\footnote{\textsc{Aut:} For details see \cite{DKL2015FooP}; \cite{KujalaDzhafarovLarsson2015};
\cite{KujalaDzhafarov2015}; \cite{deBarrosDzhafarovKujalaOas2015}.} 
\begin{enumerate}
\item \noindent set $Q$ of ``objects being measured,''\index{measurement!--- object}
\item \noindent set $C$ of ``contexts of measurements,''\index{measurement!--- context}
\item \noindent relation ``object $q$ is measured in context $c$,''\index{measurement!--- as random variable}
$q\Yleft c$, and
\item \noindent set of ``measurements,'' random variables $R=\left\{ R_{q}^{c}:q\Yleft c\right\} _{q\in Q}^{c\in C}$. 
\end{enumerate}
\noindent \ppp Two postulates of the theory are\index{Contextuality-by-Default!---, postulates}
\begin{enumerate}
\item \noindent for a given context $c\in C$, the set of measurements $R^{c}=\left\{ R_{q}^{c}:q\Yleft c\right\} _{q\in Q}$
is a random variable (which means the measurements are jointly distributed);\index{joint distribution!--- within a context}
\item any two measurements $R_{q}^{c},R_{q'}^{c'}$ with $c\not=c'$ are
stochastically unrelated (whether $q=q'$ or not).\index{joint distribution!--- not defined}\index{stochastically unrelated}
\end{enumerate}
\noindent \ppp We call the set $R_{q}=\left\{ R_{q}^{c}:q\Yleft c\right\} ^{c\in C}$
a \index{connection}\emph{connection} for object $q$. The elements of a connection
are pairwise stochastically unrelated. Let $T_{q}=\left\{ T_{q}^{c}:q\Yleft c\right\} ^{c\in C}$
be a coupling for connection $R_{q}$, i.e., $T_{q}^{c}$ has the
same distribution as $R_{q}^{c}$ for all $c$ such that $q\Yleft c$.
This coupling is called \index{coupling!---, maximal}\emph{maximal} if $\Pr\left[\textnormal{for any }c,c'\in C,\;T_{q}^{c}=T_{q}^{c'}\right]$
is maximal across all possible couplings for $R_{q}$. Such a coupling
exist, and the \emph{maximal probability} in question, $p_{q}$, is
uniquely defined. 

\noindent \ppp Let $S=\left\{ S_{q}^{c}:q\Yleft c\right\} _{q\in Q}^{c\in C}$
be a coupling for the system $R$, i.e., $S^{c}=\left\{ S_{q}^{c}:q\Yleft c\right\} _{q\in Q}$
is distributed as $R^{c}=\left\{ R_{q}^{c}:q\Yleft c\right\} _{q\in Q}$
for all $q\Yleft c$. This coupling is called \index{coupling!---, maximally connected}\emph{maximally connected}
if 
\[
\Pr\left[\textnormal{for any }c,c'\in C,\;S_{q}^{c}=S_{q}^{c'}\right]=p_{q}
\]
 for every $q\in Q$. If $R$ has a maximally connected coupling it
is \index{system of measurements!---, (non)contextual}\emph{noncontextual}. If $R$ does not have a maximally connected
coupling it is \index{system of measurements!---, (non)contextual}\emph{contextual}.

\noindent \qq Do we know criteria of (non)contextuality analogous
to the one you mentioned before, for the system $\left\{ \left(R_{A}^{V},R_{B}^{V}\right),\left(R_{A}^{W},R_{B}^{W}\right)\right\} $?

\noindent \pp Yes, this was a special (in fact, simplest) example
of a broad class of systems for which we have criteria of (non)contextuality.
These systems are called \index{system of measurements!---, cyclic}\emph{cyclic}, and they are defined as follows:
for some $n>1$,
\begin{enumerate}
\item \noindent the set of objects is $Q=\left\{ q_{1},\ldots,q_{n}\right\} $,
\item the set of contexts is $C=\left\{ c_{1},\ldots,c_{n}\right\} $,
\item for each object $q$ there are precisely two contexts $c,c'$ such
that $q\Yleft c$ and $q\Yleft c'$,
\item for each context $c$ there are precisely two objects $q,q'$ such
that $q\Yleft c$ and $q'\Yleft c$,
\item all $R_{q}^{c}$ with $q\Yleft c$ are binary, $\pm1$. 
\end{enumerate}
\ppp By appropriate enumeration we can always achieve a cyclic structure:
$q_{i}\Yleft c_{i}$ and $q_{i\oplus1}\Yleft c_{i}$ for $i=1,\ldots,n$
(where $\oplus$ is cyclic increment, with $n\oplus1=1$). The system
is noncontextual if and only if\index{system of measurements!---, cyclic!---, criterion of (non)contextuality}
\[
\max_{\substack{\textnormal{odd number}\\
\textnormal{of minuses}
}
}\sum_{i=1}^{n}\left(\pm\left\langle R_{i}^{i}R_{i\oplus1}^{i}\right\rangle \right)\leq\left(n-2\right)+\sum_{i=1}^{n}\left|\left\langle R_{i\oplus1}^{i}\right\rangle -\left\langle R_{i\oplus1}^{i\oplus1}\right\rangle \right|,
\]
where each $\pm\left\langle R_{i}^{i}R_{i\oplus1}^{i}\right\rangle $
is replaced with $+\left\langle R_{i}^{i}R_{i\oplus1}^{i}\right\rangle $
or $-\left\langle R_{i}^{i}R_{i\oplus1}^{i}\right\rangle $ so that
the minus is chosen an odd number of times ($1,3,\ldots$). 

\noindent \qq This looks like a good point at which to adjourn, my
dear Expositor. I have to think this all over. I am sure I will come
up with more questions and misgivings.

\noindent \pp I am sure you will, my dear Interlocutor. Until then,
good bye.\footnote{\textsc{Aut:} For a brief overview of CbD, see \cite{DKLC2015}.}

\subsection*{Acknowledgments}

This work was supported by NSF grant SES-1155956 and AFOSR grant FA9550-14-1-0318.
We thank Lasse Leskel{\"a} of Aalto University, Matt Jones of University
of Colorado, Victor H. Cervantes of Purdue, and Ru Zhang of Purdue for most helpful discussions.

\bibliographystyle{apalike}
\bibliography{si.bib}

\end{document}